\documentclass[aps,pra,twocolumn,superscriptaddress,showpacs,floatfix]{revtex4-1}

\usepackage{times,amssymb,amsmath}
\usepackage{array}
\usepackage{ifpdf}
\usepackage[varg]{txfonts} 
\usepackage{graphicx}
\renewcommand{\vec}[1]{\mathbf{#1}}

\mathchardef\mhyphen="2D

\renewcommand{\Re}{\text{Re}\,}

\renewcommand{\Im}{\text{Im}\,}
\usepackage{cleveref}
\newtheorem{theorem}{Theorem}
\newtheorem{corollary}{Corollary}
\begin{document}


\title{Time-evolution of excitations in  normal Fermi liquids}
\author{Y. Pavlyukh}
\email[]{yaroslav.pavlyukh@physik.uni-halle.de}
\affiliation{Institut f\"{u}r Physik, Martin-Luther-Universit\"{a}t
  Halle-Wittenberg, 06120 Halle, Germany}
\author{A. Rubio}
\affiliation{Nano-Bio Spectroscopy Group and ETSF Scientific Development Centre,
 Dpto. de F{\'i}sica de Materiales, Universidad del Pa{\'i}s Vasco,
 CFM CSIC-UPV/EHU-MPC and DIPC, Av. Tolosa 72, E-20018 San Sebasti{\'a}n, Spain}
\author{J. Berakdar}
\affiliation{Institut f\"{u}r Physik, Martin-Luther-Universit\"{a}t
  Halle-Wittenberg, 06120 Halle, Germany}\date{\today}
\begin{abstract}
We inspect the initial and the long time evolution of excitations a Fermi liquids by
analyzing the time behavior of the electron spectral function.  Focusing on the short-time
limit we study the electron-boson model for the homogenous electron gas and apply the
first order (in boson propagator) cumulant expansion of the electron Green's function.  In
addition to a quadratic decay in time upon triggering the excitation, we identify
non-analytic terms in the time expansion similar to those found in the Fermi edge
singularity phenomenon.  We also demonstrate that the exponential decay in time in the
long-time limit is inconsistent with the GW approximation for the self-energy. The
background for this is the Paley-Wiener theorem of complex analysis. To reconcile with the
Fermi liquid behavior an inclusion of higher order diagrams (in the screened Coulomb
interaction) is required.
\end{abstract}
\pacs{71.10.-w,31.15.A-,73.22.Dj}
\maketitle
\section{Introduction}
Recently, short-time dynamics of excitations in a Fermi liquid has gained an increased
attention due to the feasibility of new spectroscopic techniques
~\cite{krausz_attosecond_2009,cavalieri_attosecond_2007,schultze_delay_2010,ott_quantum_2012}
capable of accessing the attosecond time regime.  For example
in~\cite{cavalieri_attosecond_2007} a time delay in the range of 80 attoseconds between
photoemission from the core-level of tungsten and from its conduction band has been
measured. A further attosecond technique relies on the initial excitation of the system by
an attosecond pulse and then tracing the excitation evolution by monitoring the response
to a second phase-locked laser pulse~\cite{ott_quantum_2012}. This delivers a view on how
quasiparticle states develop and decay in time on a scale well below the time that one may
extract from their spectral width.  The existent experiments on the attosecond time delay
in photoemission call not only for a theoretical determination of the measured time delays
but most importantly pose a question of what new physics can be gained from these highly
sophisticated experiments. Does the spectral width of the QP peak encompass the complete
information on how the QP is born/decays?  As shown below, this is indeed not the
case. The initial stage of the QP evolution follows a different time law as in the long
time limit and the decay constants have a specific and materials dependent nature,
endorsing thus the novel physics that can be gained with the attosecond metrology.
 
 Theoretically, several methods can be used to explore the system's dynamics numerically:
the density matrix approach, the time-dependent density functional or density matrix
renormalization group theories and the non-equilibrium Green's
functions~\cite{thygesen_nonequilibrium_2007, dahlen_solving_2007,
thygesen_conserving_2008, pal_conserving_2009, pal_optical_2011,
uimonen_comparative_2011}. For the latter approach, which can be written in the form the
Kadanoff-Baym equations, the two-times lesser (greater) Green's functions $\vec
G^{\lessgtr}(t_1,t_2)$ defined on the Keldysh time-loop contour are the central
quantities. In the equilibrium situation these functions are directly related via the
so-called Kubo-Martin-Schwinger boundary conditions to the advanced and retarded Green's
functions multiplied by the corresponding electron/hole distribution functions. In the
nonequilibrium case they are complicated two-times quantities which can be found only
approximately by using some approximations for the electron
self-energy~\cite{kwong_real-time_2000,ness_gw_2011}. In order to judge the accuracy of
such approximations it is desirable to have an analytic solution for some limiting cases.

On the basis of spectral moments calculations~\cite{vogt_spectral_2004} for a 3D electron
gas we conjectured an interpolative form of the electron spectral function at the energy
$\epsilon$:
\begin{equation}
A(t;\epsilon)=A_{QP}(t;\epsilon)\,\exp\left(-\gamma(\epsilon) \frac{t^2}{t+\tau(\epsilon)}\right)\label{eq:At},
\end{equation}
with $A_{QP}(t;\epsilon)$ the oscillatory quasiparticle part.  At longer times
$A(t;\epsilon)$ decays exponentially as expected from very general considerations based on
the Landau theory of Fermi liquids and corrects a spurious divergence of the second
spectral moment. The latter can be traced back to the fact that the exponential decay requires
a certain time ($\tau(\epsilon)$) to set in and that initially the decay is quadratic:
\begin{equation}
\frac{d}{dt}A(t;\epsilon)\xrightarrow{t\rightarrow0}
-\sigma^2(\epsilon)\,t\label{eq:shr-tm}
\end{equation}
The spectral function, being essentially the decisive \emph{equilibrium} property, also
enters the non-equilibrium two-times dynamics via the \emph{generalized Kadanoff-Baym
Ansatz}~\cite{lipavsky_generalized_1986,kita_introduction_2010}:
\begin{equation}
-i\vec{G}^{\lessgtr}(t_1,t_2)=\vec{G}^{r}(t_1-t_2)\vec G^{\lessgtr}(t_2)-
\vec G^{\lessgtr}(t_1)\vec G^{a}(t_1-t_2),\label{eq:GKBA}
\end{equation}
and, thus, can be used to calibrate the approximate numerical solutions. Although being
quite natural, our underlying assumptions leading to \eqref{eq:At} need to be rigorously
verified, a task tackled here.  In addition, it is desirable to quantify the parameters
appearing in \eqref{eq:At} in terms of measurable physical quantities and to offer a
scheme for their computations. These are the goals of the present work.

The cumulant expansion is a well established procedure to study the dynamics of many-body
systems in the time domain. It amounts to writing the electron Green's function in the
form:
\begin{equation}
\mathcal G(k,t)=\mathcal{G}^0(k,t)\,e^{C(k,t)},\label{eq:gcumulant}
\end{equation}
The approach gained its wide recognition after Nozi{\`e}res and de
Dominicis~\cite{nozieres_singularities_1969} demonstrated an exact solution of a complex
integral equation for the cumulant function $C(k,t)$ for the Fermi edge singularity model.
Initial applications to model systems include the Mahan's
treatment~\cite{mahan_phonon-broadened_1966} of the Fr{\"o}hlich Hamiltonian, the
Langreth's study of the singularities in the X-ray spectra of
metals~\cite{langreth_singularities_1970}, or the fourth-order cumulant expansion for the
Holstein model by Gunnarsson \emph{et al}.~\cite{gunnarsson_corrections_1994}. More
recently the approach was also applied, among others, to describe phenomena in realistic
systems such as multiple plasmon satellites in Na and Al spectral
functions~\cite{aryasetiawan_multiple_1996} or in the valence photoemission of
semiconductors~\cite{guzzo_valence_2011}. Also similarity of this method to the
coupled-cluster method broadly used in quantum chemistry is well
known~\cite{schork_derivation_1992}.

Typically the method is applied to systems which allow a distinct separation of the
Hamiltonian into the parts allowing for the analytical treatment and a coupling that needs
to be treated perturbatively. A generic example is provided by the electron-boson
Hamiltonian describing a fermionic subsystem (the usual quasiparticles) interacting with
the bosonic excitations (e.g. phonon or the density fluctuations as will be considered
below):
\begin{equation}
\mathcal H=\sum_k\epsilon_k c_k^\dagger c_k+\sum_q\omega_qb_q^\dagger b_q
+\sum_{k,k'}\sum_{q}\mathcal V^q_{kk'}(b_q^\dagger+b_q)c_k^\dagger c_{k'},~\label{eq:Hmodel}
\end{equation}
where we narrowed the domain of quantum numbers characterizing the system to a single
wave-vector $\vec k$ as in the case of the homogeneous electron gas model. However, the
formalism can easily be extended to realistic systems.  Here $\omega_q$ describes the
energies of bosonic excitations, $\epsilon_k=k^2/2$ is the usual particle dispersion in a
weakly interacting Fermi liquid and $V^{q}_{k,k'}$ is the coupling potential.  We will use
the concept of long-lived fermionic excitations (quasiparticles) as a defining property of
the normal Fermi liquid state~\cite{giuliani_quantum_2005}. In fact this requirement is
quite restrictive as it breaks down in, e.g., low-dimensional systems.

 Under some circumstances the model \eqref{eq:Hmodel} is exact: typically this is the case
when certain matrix elements of the Coulomb interactions between a test particle (such as
a deep core hole~\cite{langreth_singularities_1970} or a high-energy
photoelectron~\cite{hedin_transition_1998}) are vanishingly small. For a more general
scenario, e.g. as we consider here, the accurateness of \eqref{eq:Hmodel} is less
obvious~\cite{guzzo_valence_2011}. In view of this fact it is interesting to consider
the connections with other theories. Parallels between the cumulant expansion and the
many-body perturbation theories (MBPT) in terms of the electron self-energy were explored by
Aryasetiawan~\cite{aryasetiawan_multiple_1996} in the lowest order. But does this
correspondence hold at an arbitrary order? Development of the formalism and an answer to
this question will be provided in Sec.~II.

The cumulant function $C(k,t)$, we show, can be written very accurately in terms of the
dynamical structure factor $\mathcal S(k,\omega)$. This allows us to obtain exact
analytical results for the prefactor $\sigma^2(\epsilon)$ of the quadratic
decay~\eqref{eq:shr-tm}. It is interesting that also terms proportional to $(-it)^3$ can be
written in a concise analytical form. These results do not require any further
approximations in addition to the ones discussed in Sec.~II and follow  from exactly
known sum-rules and the asymptotic behavior of $\mathcal S(k,\omega)$ (Sec.~III).

The quasiparticle uncertainty $\sigma^2(\epsilon)$~\eqref{eq:shr-tm} can alternatively be
obtained from the zeroth spectral moment of the electron
self-energy~\cite{pavlyukh_initial_2011}. While this approach is perfectly justified for
finite systems~\cite{pavlyukh_communication:_2011} a careful analysis must be done in the
case of the homogenous electron gas (HEG) model. Here the difficulties arise from a
particular asymptotic behavior of the electron self-energy: decay as $\omega^{-3/2}$ for
$\omega\rightarrow\infty$ and its vanishing imaginary part for $\omega<\omega^*(k)$.  In
Sec.~IV we trace the origins of these features. On the basis of the Paley-Wiener (PW)
theorem~\cite{paley_fourier_1934} we demonstrate that such restriction of the spectrum
implies an unphysical spectral function which in the time domain asymptotically decays
faster than the exponent. This is, however, a consequence of GW approximation for
$\Sigma(k,\omega)$. The paradox is further resolved by considering higher order
contributions to the electron self-energy.
\section{Cumulant vs. self-energy expansion}
For the Hamiltonian \eqref{eq:Hmodel} the interaction between fermionic and bosonic
degrees of freedom is given by the fluctuation potential~\cite{hedin_transition_1998} with
matrix elements:
\begin{equation}
\mathcal V^{q}_{k,k'}=\frac{4\pi}{|\vec k-\vec k'|^2}\,\rho^q_{\vec k-\vec k'},
\end{equation}
where $q$ is a quantum number characterizing the excited bosonic states and $\rho^q_{\vec
  k}$ is the $\vec k$th Fourier component of the fluctuation density operator between the
  ground state and a state with one boson with the quantum number $q$ excited. The choice
  of the interaction form is not arbitrary: it guarantees that the lowest order diagram
  for $\mathcal G(k,t)$ in the model \eqref{eq:Hmodel} corresponds to the GW approximation
  (for more details on this approximation see Sec.~\ref{sec:gw}) for the initial fermionic
  system.

It is instructive to derive $C(k,t)$ starting from MBPT and using the method of
Aryasetiawan~\cite{aryasetiawan_multiple_1996}. By comparing the expansion of the
exponential in \eqref{eq:gcumulant} to that obtained by iterating the Dyson's equation:
\begin{multline}
\mathcal G(k,t)=\mathcal G^0(k,t)\\+\int\!\!d\tau \!\!\int\!\! d\tau'
 \mathcal G^0(k,t-\tau)\,\Sigma(k,\tau-\tau')\, \mathcal G(k,\tau')\label{eq:dyson}
\end{multline}
we can easily verify that $C(k,t)$ and $\Sigma(k,t)$ should have the same lowest order
expression in terms of the interaction.  We separately consider the particle ($k>k_F$) and
the hole ($k<k_F$) cases. The non-interacting Green's function is given by $\mathcal
G_+^0(k,t)=-i\theta(t)\,e^{-i\epsilon_kt}$ and $\mathcal
G_-^0(k,t)=i\theta(-t)\,e^{-i\epsilon_kt}$, respectively. We represent the full Green's
functions as $\mathcal G_\pm(k,t)=\mathcal G_\pm^0(k,t)\,e^{C_\pm(t)}$. These notations
are different from the ones used by Langreth where they denoted two differently defined
Green's functions. For the electron self-energy we adopt the standard (non
self-consistent) expression (cf. Eq.~(25.1) of \cite{hedin_effects_1970}):
\[
\Sigma(k,t)=i\int \frac{d\vec q}{(2\pi)^3}\,\mathcal W(\vec q-\vec k,t+\delta)\,\mathcal
G^0(\vec q,t),\quad \delta\rightarrow+0,
\]
with the screened Coulomb interaction given by:
\begin{equation}
\mathcal W(k,\omega)=v(k)+v^2\Pi(k,\omega)=
v(k)+\sum_q\frac{2\omega_q \big|\,V^{q}_k|^2}{\omega^2-\omega_q^2},
\end{equation}
where $v(k)=4\pi/k^2$ is the Coulomb potential and $\Pi(k,\omega)$ is the full bosonic
propagator or the \emph{density-density response function} in this particular case.  The
latter is related by the \emph{fluctuation-dissipation theorem} (cf. Eq.~3.74 of
\cite{giuliani_quantum_2005}) to the \emph{dynamical structure factor}:
\[
\mathcal S(k,\omega)=-\frac1\pi\Im \Pi(k,\omega).
\]
For the $\omega$-dependent part of the screened Coulomb interaction we use the spectral
representation:
\[
\widetilde{\mathcal W}(\vec k,t)=i v^2(k)\int_0^\infty \!\!d\omega\, \mathcal S(k,\omega)\,e^{-i\omega t},
\]
where the dynamic structure factor is expressed in terms of the imaginary part of the
dielectric function ($\varepsilon=\varepsilon'+i\varepsilon''$):
\[
\mathcal S(k,\omega)=\frac{k^2}{4\pi^2}\frac{\varepsilon''(k,\omega)}
{|\varepsilon(k,\omega)|^2}\,\theta(\omega).
\]
Clearly, four cases arise depending on the length of $\vec q$ and $\vec k$.
\paragraph{$k>k_F$ and $q>k_F$:}
Inserting the expressions for $\Sigma$ and $\widetilde{\mathcal W}$ in Eq.~\eqref{eq:dyson} we
obtain:
\begin{multline}
C_+(k,t)=-\int_{q>k_F}\!\frac{d\vec q}{(2\pi)^3}v^2(q)\int_0^\infty\!d\omega\, \mathcal
S(\vec q-\vec k,\omega)\,e^{-i\omega\delta}\\
\times\iint_{D_+}d(\tau\tau')\,e^{i(\epsilon_k-\epsilon_q-\omega)(\tau-\tau')},\label{eq:c+0}
\end{multline}
where the integration domain is determined by the condition
$\theta(t)\,\theta(t-\tau)\,\theta(\tau')\,\theta(\tau-\tau')$.
This is, in fact, a finite domain which can be integrated as follows:
\begin{multline}
f(\nu)\equiv\iint_{D_+}d(\tau\tau')\,e^{i\nu(\tau-\tau')}
=\int_0^td\tau\int_0^\tau\!d\tau'\,e^{i\nu(\tau-\tau')}\\
=\int_0^t d\tau(t-\tau)\,e^{i\nu\tau}=\frac{1+i\nu  t-e^{i\nu t}}{\nu^2}.
\label{eq:fdef}
\end{multline}
\paragraph{$k>k_F$ and $q<k_F$:} We have to use the hole propagator for the intermediate
line. This changes the sign of the expression and modifies the integration domain which we
represent as two terms:
\begin{equation}
\theta(t-\tau)\,\theta(\tau')\,\theta(\tau'-\tau)=
\theta(t-\tau)\,\theta(\tau')\,\big(1-\theta(\tau-\tau')\big)\label{eq:Dmod}
\end{equation}
The first term here represents an additional contribution pertinent to GW approximation
only:
\begin{equation}
\int_{-\infty}^td\tau\int_0^\infty\!d\tau'\,e^{i\nu(\tau-\tau')}= -\frac{e^{i\nu t}}{\nu^2}.
\label{eq:cgw}
\end{equation}
For the second term in \eqref{eq:Dmod} we have the same form and sign as in
\eqref{eq:c+0}.  Thus, the integration over $q<k_F$ can be combined with \eqref{eq:c+0}
resulting in the sum over all momenta.

\paragraph{$k<k_F$ and $q<k_F$:} The procedure is similar with the only difference in the domain of the integration
$D_-=\theta(-t)\,\theta(\tau-t)\,\theta(-\tau')\,\theta(\tau'-\tau)$ which also integrates
in terms of  $f(\nu)$:
\[
\iint_{D_-}d(\tau\tau')\,e^{i\nu(\tau-\tau')}
=\int_t^0d\tau\int_\tau^0\!d\tau'\,e^{i\nu(\tau-\tau')}=f(\nu).
\]
Hence, $C_-(t)$ can be written in the same form as \eqref{eq:c+0} and we will use $C(k,t)$
as a common symbol for both $C_\pm(k,t)$.  There also is a contribution from the
intermediate hole line $q>k_F$ to $C_-(t)$ which can be evaluated along the same lines as
\eqref{eq:cgw}.  Finally, we redefine the variable for momentum integration as $\vec
q-\vec k\rightarrow \vec q$ and obtain in line with
Langreth~\cite{langreth_singularities_1970}:
\begin{multline}
\label{eq:cgen}
C(k,t)=-\sum_q v^2(q)\int_{0}^{\infty}\!d\omega\,\,\mathcal S(q,\omega)\\
\times f\big(\epsilon_{|\vec k|}-\epsilon_{|\vec k+\vec q|}-\omega,t\big).
\end{multline}

The central quantity of this study -- the dynamical structure factor -- although expressed
almost identically (except for the \eqref{eq:cgw} terms) in the many-body perturbation and
in the cumulant expansion theories, originates from different approximations. In the
former case it is the vertex function in the expression for the self-energy that is
neglected, while for the latter it is assumed that the Hamiltonian can be written in the
electron-boson form \eqref{eq:Hmodel}. For the homogenous electron gas model the
justification mostly comes from MBPT although $\mathcal S(k,\omega)$ can be rather
complicated function even for simple systems~\cite{takada_dynamical_2002}.
\section{Short and long time limits\label{sec:limits}}
Equation \eqref{eq:cgen} is general enough to treat all the cases presented in
Tab.~\ref{tab:models}. Compared to Eq.~(44) of \cite{langreth_singularities_1970} we
additionally allow the test particle to scatter (i.e. to exchange its momentum with the
bosonic excitations) whereas a deep core in \cite{langreth_singularities_1970} is assumed
to have an infinite effective mass.  The short-time limit of the electron Green's function
crucially depends on the exact form of fermionic dispersion, on the boundness of the
bosonic spectrum and on the actual form of the coupling potential.
 \begin{table*}
 \caption{Electron-boson models and main results for the long and short-time limits of the
   electron Green's function\label{tab:models}}
 \begin{ruledtabular}
 \begin{tabular}{c|ccc}
Fermionic dispersion&Dispersionless phonons&Dispersionless plasmons&Electron-hole pairs\\\hline
\parbox[c][1.2cm][c]{3.5cm}{\raggedright Deep hole:
$H_0=Ec^\dagger c$}&&\parbox[c][1.2cm][c]{3.5cm}{\raggedright $\displaystyle C(t)\sim
  e^{-i\omega_pt}$\\Langreth (1970)}
&\parbox[c][1.2cm][c]{4cm}{\raggedright$\displaystyle C(t)
\sim-\alpha\big[\ln|Dt|\mp i\pi/2\,\mathrm{sgn}{t}\big]$\\Langreth (1970)}\\
\parbox[c][1.2cm][c]{3.5cm}{\raggedright Valence states:
$\displaystyle H_0=\epsilon_kc_k^\dagger c_k$}&\parbox[c][1.2cm][c]{3.5cm}{\raggedright$\displaystyle
  C(t)\sim\frac1{\omega_0}(2it)^{1/2}\,e^{-i\omega_0  t}$\\ Mahan (1961) }&
\parbox[c][1.2cm][c]{3.5cm}{\raggedright$\displaystyle C(t)\sim e^{-i\omega_pt}$\\
  Aryasetiawan (1996)}& this work
 \end{tabular}
 \end{ruledtabular}
 \end{table*}

As a first application we compute the leading expansion coefficients of the cumulant function
in the short-time limit:
\begin{equation}
C(k,t)=-\frac{\sigma^2(k)}{2!}t^2+\frac{c_3(k)}{3!}t^3+\ldots
\label{eq:expan}
\end{equation}
We note, however, that such expansion does not imply analyticity of the function in
vicinity of $t=0$. Just the opposite, higher expansion coefficients diverge starting from
$c_6$ in 2d case and from $c_7$ in 3d based on very general properties of the
density-density response function ($\Im \Pi(k,\omega)\sim\omega^{-4-d/2}$, p.~139 of
\cite{giuliani_quantum_2005}).  Although where  the divergence occurs exactly can be
modified by including higher order terms (in $\mathcal S(k,\omega)$) in the expression for
the cumulant function, this will not restore the analyticity. The prefactor of the
quadratic decay (Eq.~\eqref{eq:shr-tm}) can be computed by evaluating the second
derivative of \eqref{eq:cgen} at $t=0$:
\begin{equation}
\sigma^2=n\sum_q v^2(q)\,\mathcal S(q),\label{eq:sigma}
\end{equation}
where the \emph{static structure factor} is defined as
\begin{equation}
\mathcal S(q)=\frac1n\int_0^\infty\!\!d\omega\, \mathcal S(q,w)
\xrightarrow{q\rightarrow 0}\frac{q^2}{2\omega_p(q)}.\label{eq:sq}
\end{equation}
It follows then that $\sigma^2$ is independent of $k$ and coincides with the local
contribution to the zeroth spectral moment of the electron self-energy obtained by Vogt
\emph{et al.}~\cite{vogt_spectral_2004}.  The long wave-length expression in
Eq.~\eqref{eq:sq} follows from the exactness of the random phase approximation (RPA) in
this limit. In the opposite case (i.e. $q\rightarrow \infty$) the structure factor
approaches unity, however, the subleading term RPA fails to reproduce. In order to
accurately compute $\sigma^2$ the parameterized structure factor of Gori-Giorgi \emph{et
al.}~\cite{Gori-Giorgi_analytic_2000} based on the quantum Monte Carlo results was
used~\cite{vogt_spectral_2004}. The convergence of the integral \eqref{eq:sigma} is
ensured by the limit:
\begin{equation}
\lim_{q\rightarrow\infty}q^{z+1}\big[\mathcal S(q)-1\big]=-\pi 2^z n g(0),
\label{eq:sqlim}
\end{equation}
where $g(0)$ is the value of the pair correlation function for two electron at the same
position and $z$ is the dimensionality of a system.

It is not obvious from the outset that the $c_3(k)$ coefficient should take a finite value: this
heavily relies on the exact form of the structure factor in the asymptotic
($q\rightarrow\infty$) limit. By using the $f$-sum rule:
\begin{equation}
\int_0^\infty\!d\omega\,\omega\,\mathcal S(q,\omega)=n\epsilon_q,
\end{equation}
where $n$ is the electron density we obtain:
\begin{equation}
c_3(k)=-in\sum_qv^2(q)
\Big[\big(\epsilon_{|\vec k|}-\epsilon_{|\vec k+\vec q|}\big)\,\mathcal S(q)+\epsilon_q\Big].
\end{equation}
In the simplest case of a hole state at the band's bottom the convergence of the integral
regardless of the system's dimension ($z$) is guaranteed by the limit \eqref{eq:sqlim}.
For $k>0$ the term linear in $k$ vanishes after the angular integration and we finally
obtain the $k$-independent result:
\[
c_3=in\sum_qv^2(q)\,\epsilon_q\,\Big[\mathcal S(q)-1\Big].
\]

Finally we notice that  the leading terms of Eq.~\eqref{eq:cgen} in the long time-limit are the
constant and the linear ones, i.e., $C(k,t)\xrightarrow{t\rightarrow\infty}
\gamma-i\,\Sigma(k,\epsilon_k)t$ as expected from the exponential quasiparticle decay
(cf. Eq.~(7) of \cite{aryasetiawan_multiple_1996}).

\paragraph{Non-analyticity of the spectral function at $t=0$}
The asymptotic behavior of the density-density response function at large $\omega$ leads
to diverging expansion coefficients in Eq.~\eqref{eq:expan}. This, in turn, gives us a
hint that the cumulant function is probably \emph{nonholomorphic} at $t=0$. Such property
is, however, not an exception, but rather the rule as Tab.~\ref{tab:models}
demonstrates. We will sketch below how all the results presented in this table can be
obtained in a unified way from Eq.~\eqref{eq:cgen} and will also show that the same
applies to the normal Fermi liquids, in particular due to the scattering of valence
electrons with the generation of electron-hole pairs (viz. ``this work'' in
Tab. ~\ref{tab:models}).

One of the most interesting scenarios is the case of a core hole coupled to electron-hole
excitations. In the limit of infinite mass of the fermion and $t\rightarrow\infty$ there
is a singular term that arises from the frequency integration in Eq.~\eqref{eq:cgen}:
\[
C(t)=-\sum_q \frac{v^2(q)}{|\varepsilon(q,0)|^2} \int_0^\infty\!d\omega\, \frac{\omega}{q}
\frac{1-e^{-i\omega t}}{\omega^2}\sim -\eta\big[\ln|Dt|+ \frac{i\pi}{2}\,\mathrm{sgn} t\big]
\]
where $0<\eta<\frac12$ is called the Anderson singularity
index~\cite{anderson_infrared_1967} which can be given in terms of the scattering phase
shifts $\delta_l$ of the statically screened potential ($\mathcal W(q,0)$) as
\[
\eta=2\sum_{l}(2l+1)\left(\frac{\delta_l}{\pi}\right)^2.
\]
In the frequency domain the resulting spectral function exhibits a singularity
$\frac{2\pi}{\Gamma(\eta)}\frac{\theta(-\omega)}{\omega^{1-\eta}}$ which for the finite
hole's mass and $z>1$ is completely washed out by the effect of scatterer recoil as was
demonstrated by Nozi{\`e}res~\cite{nozieres_effect_1994}. This equivalently can be seen
from our model \eqref{eq:cgen} where the momentum angular integration of the function
$f\big(\epsilon_{|\vec k|}-\epsilon_{|\vec k+\vec q|}-\omega,t\big)$ removes the
singularity.

The cumulant function resulting from the interaction with plasmons has a simple structure
which likewise follows from \eqref{eq:cgen} by using the limiting form of the structure
factor:
\[
\mathcal S(q,\omega)\xrightarrow{q\rightarrow 0}
\frac{q^2}{8\pi}\frac{\omega_p(0)^2}{\omega_p(q)}\delta(\omega-\omega_p(q)).
\]
In the frequency domain this leads to the main quasiparticle peak accompanied by a
sequence of satellites displaced by
$n\omega_p(0)$~\cite{minnhagen_aspects_1975,ness_gw_2011}. In many realistic materials
these are indeed observed features~\cite{aryasetiawan_multiple_1996,guzzo_valence_2011}.
Recalling our remark in the introduction concerning the current status of the experimental
attosecond spectroscopy, it is obvious that these structures are interesting candidates
for the tracking of the development of main and satellite peaks to their static limit.

Another interesting case, likewise in the $t\rightarrow\infty$ limit, arises from the
polar coupling $\sim1/q$ between an electron and the non-dispersive optical phonon with
the energy $\omega_0$~\cite{mahan_phonon-broadened_1966}. It results in the effective
structure factor $\mathcal S(q,\omega)\sim 1/q^2\delta(\omega-\omega_0)$. There, the
non-analytic terms stem from the momentum integration:
\[
\sum_q \frac{1}{q^2} e^{-i\epsilon_q\tau}
=\frac{4\pi}{(2\pi)^3}\int_0^\infty\!dq\,e^{-i\epsilon_q
\tau}=\frac{1}{(2\pi)^{3/2}}\left(\frac{2}{i\tau}\right)^{1/2}.
\]
For the momentum state $k=0$ we can use the representation of $ f\big(\epsilon_{|\vec
  k|}-\epsilon_{|\vec k+\vec q|}-\omega,t\big)$ in terms of a single
  time-integral~\eqref{eq:fdef} and obtain:
\begin{multline*}
C(0,t)\sim\int_0^t d\tau(t-\tau)\,\left(\frac{2}{i\tau}\right)^{1/2}e^{-i\omega_0\tau}
\sim \frac1{\omega_0}(2it)^{1/2}\,e^{-i\omega_0  t}.
\end{multline*}

In the opposite case, i.~e. $t\rightarrow 0$, the non-analytic terms originate from the
coupling to the particle-hole ($p\mhyphen h$) continuum. We can split the momentum
integration into a finite interval $q<q_c$ yielding just the well-behaved analytic part of
$C(k,t)$ and the interval extending to infinity. The value of $q_c$ can always be chosen
large enough so that the real part of the dielectric function on the second interval
approaches unity. This considerably simplifies the dynamical structure factor which
results now from the imaginary part of the Lindhard formula (Eq.~(5.35) of
\cite{hedin_effects_1970} for $z=3$) only:
\[
S(q,\omega)=\frac{1}{\pi}\frac{1}{v(q)}\frac{\alpha r_s}{\tilde q^3}\left(
1-\frac14\left(\tilde q- \frac{\tilde\omega}{\tilde q}\right)^2
\right),
\]
where the tilde denotes the use of rescaled quantities, i.e. $q=\tilde q k_F$, $\omega=\tilde
\omega \epsilon_F$, $\tilde\tau=\tau/\epsilon_F$ and so on.  After the substitution
$\tilde \omega=\tilde q^2+2\tilde q\lambda$ we can first integrate over the interval
$|\lambda|<1$.  This yields a trigonometric expression which is just a constant in the
lowest order of $2\tilde q\tilde \tau$:
\[
\int d\omega\,\mathcal S(q,\omega)\,e^{-i(\tilde\omega-\tilde q^2)\tilde\tau}
=\frac{1}{3\pi^2}k_F^3+\mathcal O(2\tilde q\tilde \tau).
\]
The leading non-analytic term of the remaining momentum integral reads:
\[
\frac{1}{(2\pi)^3}\int\! q^2 dq\,v^2(q)\,  e^{-2i\tilde q^2 \tilde \tau}\sim
-\frac{2}{k_F}\left(\frac{2i\tilde\tau}{\pi}\right)^{1/2}.
\]
Finally we perform the time integration as in \eqref{eq:fdef}:
\begin{equation}
C(0,t)=\frac{2k_F^2}{3\pi^2}\int_0^t \!d\tau\,(t-\tau)
\left(\frac{2i\tilde\tau}{\pi}\right)^{1/2}=
-\frac{8}{45}(\alpha r_s)^2\left(\frac{2it}{\pi}\right)^{5/2}.
\label{eq:sqrt}
\end{equation}
Such time-dependence is easy to reconcile with well known asymptotic behavior of the
electron self-energy as a function of
frequency~\cite{bose_asymptotic_1975,vogt_spectral_2004}:
\[
\Im\tilde\Sigma(k,\tilde\omega)\xrightarrow{\omega\rightarrow\infty}=-\frac{16\sqrt{2}}{3\pi}
\frac{(\alpha r_s)^2}{\tilde \omega^{3/2}}.
\]
To see the connection we express asymptotically the spectral function as $A(k,\omega)\sim
C/\omega^{2+3/2}$ and perform the Fourier transform. Since at $\omega\rightarrow -\infty$
the spectral function decays faster, in fact on the GW level it is even zero below a
certain threshold value of $\omega$, it is sufficient to perform the transform on a
semi-bounded interval:
\[
A(k,t)\sim\int_{\omega_c}^\infty\frac{d\omega}{2\pi}\frac{C}{\omega^{7/2}}\,e^{-i\omega t}.
\]
Among several resulting terms one has to pick up the one independent of the cut-off
$\omega_c$. It exhibits the same time-dependence and the density scaling $\sim(\alpha
r_s)^2$ as Eq.~\eqref{eq:sqrt}.
\section{Self-energy in frequency space\label{sec:gw}}
In 1965 Lars Hedin formulated a system of functional equations~\cite{hedin_new_1965},
carrying by now his name, that relate the electron self-energy $\Sigma(12)$, the
irreducible polarization propagator $\mathcal P(12)$, the screened Coulomb interaction
$\mathcal W(12)$, the vertex function $\Gamma(12;3)$ and the electron Green's function
$\mathcal G(12)$. The Hedin equations are becoming one of the major theoretical tools for
the treatment of correlated many-particle
systems~\cite{onida_electronic_2002,aryasetiawan_beyond_2009,ren_resolution--identity_2012}. The
homogeneous electron gas (HEG) in two or three dimensions is a prototypic model which
allows for testing various approximations to the exact Hedin's equations. Earlier
applications revealed important features of the single-particle
spectrum~\cite{lundqvist_single-particle_1967,lundqvist_single-particle_1967-1,lundqvist_single-particle_1968}.
These single-shot calculations were extended by several authors to the self-consistent
level~\cite{von_barth_self-consistent_1996,holm_fully_1998}, higher order diagrams were
included~\cite{shirley_self-consistent_1996,takada_inclusion_2001}.

In view of large efforts devoted to the study of these model systems it is surprising that
some aspects remained unnoticed. Thus, it is commonly believed that HEG in two or three
dimensions serves as a perfect illustration of the Fermi liquid
concept~\cite{giuliani_quantum_2005}, that is a many-body fermionic systems with
long-lived excitations: \emph{quasiparticles}. Two marked properties distinguish them from
other excited states: i) they can be brought in a direct correspondence with real
particles (electrons) of a fictitious non-interacting many-body system; ii) they are
characterized by the life-time, which tends to infinity as the particle's energy
approaches the Fermi level ($\epsilon_F$). It also implies that asymptotically the decay
is exponential $\exp(-\gamma t)$, with the decay constant being quadratically dependent on
the energy ($\gamma(\epsilon)\sim \epsilon^2/\epsilon_F$). At
$\epsilon\rightarrow\epsilon_F$ this constant can be computed perturbatively, and it is
sufficient to consider the lowest-order term giving a non-vanishing imaginary part of the
self-energy. In view of this it is intriguing that a rigorous proof can be given that the
lowest-order diagram yield the spectral function inconsistent with the asymptotic
exponential decay.

\begin{figure}[t!]
\includegraphics[width=\columnwidth]{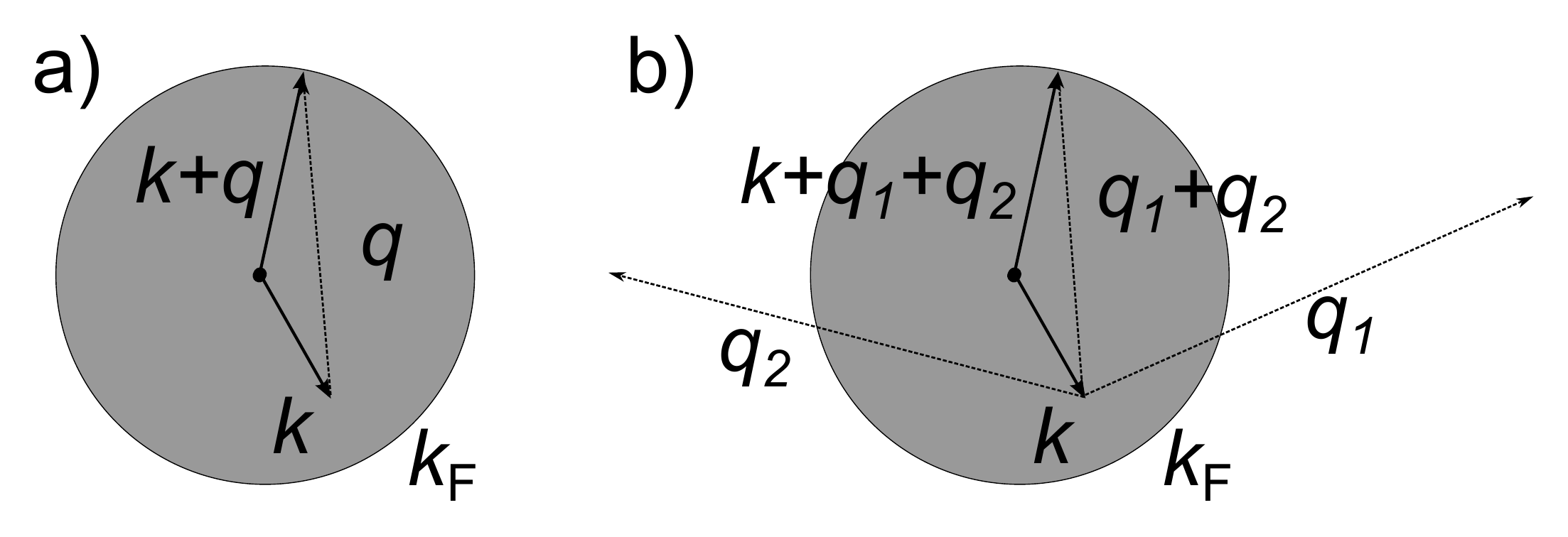}
\caption{a) First and b) second order hole scattering mechanisms. While the two
  excitations in b) carry in total the same momentum as a single excitation in a) the
  energy transfer is much larger.
\label{fig:sketch}}
\end{figure}

One recognizes that pronounced features in the spectral function appear at energies $E_k$
that are approximately given by $E_k=\epsilon_k+\Sigma(k,E_k)$, where
$\Im\Sigma(k,\omega)\sim\delta(\omega-\epsilon_{\vec{|k+q|}}\pm\omega_q)$. These
resonances are surrounded by the incoherent background which has the same extend as the
self-energy:
\begin{multline}
A(k,\omega)=\frac1{\pi}|\Im \mathcal G(k,\omega)|\\
=\frac1{\pi}\frac{|\Im \Sigma(k,\omega)|}
{|\omega-\epsilon_k-\Re\Sigma(k,\omega)|^2+|\Im \Sigma(k,\omega)|^2}.
\label{eq:Aw}
\end{multline}
In the lowest order of the screened interaction a particle can only loose its energy
($\epsilon_k$) by generating a \emph{single} bosonic excitation $\omega_q$. Since only a
finite momentum can be transfered also $\omega_q$ is finite and, thus, the self-energy has
a semi-bounded support (limited from below) (Fig. \ref{fig:sketch}). From this, in view of
\eqref{eq:Aw} follows $A(k,\omega)=0$ for $\omega<\omega^*(k)$.

The last property allows us to apply the Paley-Wiener (PW)
theorem~\cite{paley_fourier_1934} which we present below for completeness in the original
formulation:
\begin{theorem}
Let $\phi(x)$ be a real non-negative function not equivalent to zero, defined for
$-\infty<x<\infty$, and of integrable square in this range. A necessary and sufficient
condition that there should exist a real- or complex-valued function $F(x$) defined in the
same range, vanishing for $x\ge x_0$ for some number $x_0$, and such that the Fourier
transform $G(x)$ of $F(x)$ should satisfy $|G(x)| = \phi(x)$, is that
\begin{equation}
  \int_{-\infty}^{\infty}\frac{|\log\,\phi(x)|}{1+x^2}\,dx<\infty.
  \label{eq:intln}
\end{equation}
\end{theorem}
The modern formulation~\footnote{see \cite{exner_open_1985} and
  \cite{stein_princeton_2003} for a pedagogical introduction} only slightly relaxes the
  conditions on the functions:
\begin{theorem}
For $\phi(x)\in L^2(\mathbb{R})$ and $\phi(x)>0$ the integral~\eqref{eq:intln} converges
$\iff$ there is a function $F\in L^2(\mathbb{R})$ with a semi-bounded support such that
$\phi=|\mathcal{F}[F]|$ a.~e. in $\mathbb{R}$, and $\mathcal{F}$ is the Fourier-Plancherel
operator.
\end{theorem}
From here follows:
\begin{corollary}
\begin{multline*}
F(y)=0\quad\text{for}\quad y<y_0\Rightarrow
\int_{\mathbb{R}}\frac{|\log\,\phi(x)|}{1+x^2}\,
  dx<\infty\quad\Rightarrow\\
|\phi(x)|\xrightarrow{x\rightarrow\infty}\exp(-Bx^\alpha),\quad\text{where}\quad\alpha<1.
\end{multline*}
\end{corollary}
If we identify now $F$ with $A(k,\omega)$ the deviation from the exponential decay for
$A(k,t)$ follows: a result in a clear contradiction with the Fermi liquid theory. Before
proceeding with the resolution of the paradox we present our method for numerical
calculation of $\omega^*(k)$. It is sufficiently general in the sense that there is no
limitation on the dimensionality ($z$) of the system and it is not limited to the first
order expression. From now on we will only be using rescaled quantitates, tilde will be
omitted for clarity.  In order to make the model amenable for the numerics we introduce
the following representation of the screened Coulomb interaction:
\begin{equation}
\mathcal W^0(k,\omega)=\frac{v(k)}2\!\int\!d\lambda\,
\left[\frac{w(k,\lambda)}{\omega-\Omega(k,\lambda)+i\eta}
-\frac{w(k,\lambda)}{\omega+\Omega(k,\lambda)-i\eta}\right],
\label{eq:Wrepr}
\end{equation}
where $w(k,\lambda)$, $\Omega(k,\lambda)$ are some real functions that will be specified
below. Our representation takes advantage of the fact that the imaginary part of the
dielectric function and the screened Coulomb interaction is different from zero only in
the stripe area in the $\omega\mhyphen k$ plane and along the plasmonic
line~\cite{giuliani_quantum_2005}. The limits for the particle-hole continuum are given
(for $z\ge2$) by:
\begin{equation}
\max\Big\{0,\omega_{-}(k)\Big\}\le|\omega|\le\omega_{+}(k),\,\text{with}\quad
\omega_\pm(k)=k^2\pm2k,\label{eq:stripe}
\end{equation}
where it is convenient to parameterize the trajectories on the stripe~\eqref{eq:stripe} as:
\[
\Omega(k,\lambda)=k^2+2\lambda k.
\]
Thus Eq.~\eqref{eq:Wrepr} is nothing but the spectral representation (see e.~g. Eq.~4
of~\cite{von_barth_self-consistent_1996}):
\[
\mathcal  W(k,\omega)=v(k)+\int_0^\infty\frac{2\omega'B(k,\omega')}{\omega^2-\omega'^2}d\omega',
\]
The integral over $\lambda$ is to be understood in a generalized sense: this parameter can
assume both discrete values when we describe a single excitation such as plasmon or be a
continuous variable for particle-hole excitations. In the former case it reduces to the
plasmon model approximation (cf. Eq.~25.11 of~\cite{hedin_effects_1970}):
\[
\mathcal  W^0(k,\omega)=v(k)\left[1+\frac{\omega_p^2(0)}{\omega^2-\omega_p^2(k)}\right].
\]
The bare Coulomb part can also be obtained from \eqref{eq:Wrepr}: consider the limit
$\Omega(k,\lambda)\rightarrow w(k,\lambda) \rightarrow\infty$. The fact that we can
represent all contributions to $W(k,\omega)$ in a unified way is crucial for our
discussion: one does not need to separately consider diagrams with bare or renormalized
interaction lines. The former can be obtained from the general case by formally taking the
limit of the final expression.

We write the Green's function as:
\[
\mathcal G^0(k,\omega)=\frac{n_k}{\omega-\epsilon_k-i\eta}+\frac{1-n_k}{\omega-\epsilon_k+i\eta},
\]
where $n_k$ denotes the occupation of the state with the momentum $k$ and consider the two
lowest order diagrams for the electron self-energy $\Sigma[\mathcal G^0,\mathcal W^0]$:
\begin{subequations}
\label{eq:Sgm}
\begin{eqnarray}
\Sigma^{(1)}(1,2)&=&i\,\mathcal G^0(1,2)\,\mathcal W^0(1^+,2),\\
\Sigma^{(2)}(1,2)&=&i^2\!\iint \mathcal W^0(1^+,4)\,\mathcal G^0(1,3)\,
\mathcal G^0(3,4)\nonumber\\&&
\quad\quad\quad\quad\times\, \mathcal G^0(4,2)\,\mathcal W^0(3^+,2)\,d(34).
\end{eqnarray}
\end{subequations}
Our representation of the screened Coulomb interaction allows to compute the electronic
self-energy relatively easy using the {\sc maple} computer algebra system~\footnote{see
Supplemental Material}. The final results can be recasted in the form of momentum and
$\lambda$ integrals over complicated domains that we denote as $\mathcal D_\pm^{(i)}$,
where $\pm$ designates particle (hole) state, and $(i)$ is the order of a diagram. Since
we are only interested in the phase-space where each diagram contributes we skip here the
explicit expressions for $\Sigma(k,\omega)$ and present only the results for $\mathcal
D_\pm^{(i)}(k,\omega)$:
\begin{subequations}
\label{eq:Dom}
\begin{eqnarray}
\mathcal D_{-}^{(1)}&=&n_p\,\delta\big(\omega-\epsilon_p+\Omega(q,\lambda)\big),\\
\mathcal D_{-}^{(2a)}&=&(1-n_{p_0})\,n_{p_1}n_{p_2}
\delta\big(\omega+ \epsilon_{p_0}-\epsilon_{p_1}-\epsilon_{p_2}\big),\\
\mathcal D_{-}^{(2b)}&=&n_{p_0}\delta\big(\omega-\epsilon_{p_0}
+\Omega(q_1,\lambda_1)+\Omega(q_2,\lambda_2)\big),\label{eq:D2b}\\
\mathcal D_{-}^{(2c)}&=&n_{p_1}\delta
\big(\omega-\epsilon_{p_1}+\Omega(q_1,\lambda_1)\big)\nonumber\\
&&\times\,\Big[1+f_a(q_1,q_2,p_1,p_2,p_0)\big(1-n_{p_1}\big)\nonumber\\
&&f_b(q_1,q_2,p_1,p_2,p_0)\big(1-n_{p_0}\big)\Big]+(1\leftrightarrow2).
\end{eqnarray}
\end{subequations}
where we introduced the following vectors $\vec p=\vec k+\vec q$, $\vec p_0=\vec k-\vec
q_1-\vec q_2$, $\vec p_1=\vec k-\vec q_1$, and $\vec p_2=\vec k-\vec q_2$. $\mathcal
D_{-}^{(1)}(k,\omega)$ describes the simplest first order process when a hole scatters to
another hole-state hereby generating a plasmon or a $p\mhyphen h$ pair. $\mathcal
D_{-}^{(2a)}$ describes a hole scattered to a two-holes-one-particle ($2h\mhyphen p$)
state, whereas $\mathcal D_{-}^{(2b)}$ stands for a process when a hole looses its energy
by the generation of two bosonic excitations. The last term is not interesting because it
just renormalizes $\mathcal D_{-}^{(1)}(k,\omega)$. Expressions for the particle states
analogous to \eqref{eq:Dom} can be obtained by the use of the particle-hole symmetry.

We compute the integrals involving $\mathcal D_\pm^{(i)}(k,\omega)$ by using the
Monte-Carlo approach and formulas for the momentum integration presented in
Appendix~\ref{A:1} (Fig.~\ref{fig:Psgm}). Thus, the first order contribution is obtained
by throwing a quartet of random numbers $(k,\,\lambda,\,q,\,y)$ consistent with the
integration domain. Henceforth, we verify the condition imposed by the $\delta$-function
and determine possible values of $\omega$.  As for $\mathcal D_{-}^{(2a)}$ the probability
distribution is obtained from a set of 5 numbers $(k,\,y_{1,2},\,q,\,Q)$, whereas we need
to additionally sample over $\lambda_{1,2}$ random variables for $\mathcal D_{-}^{(2b)}$.
\begin{figure}[t!]
\includegraphics[width=\columnwidth]{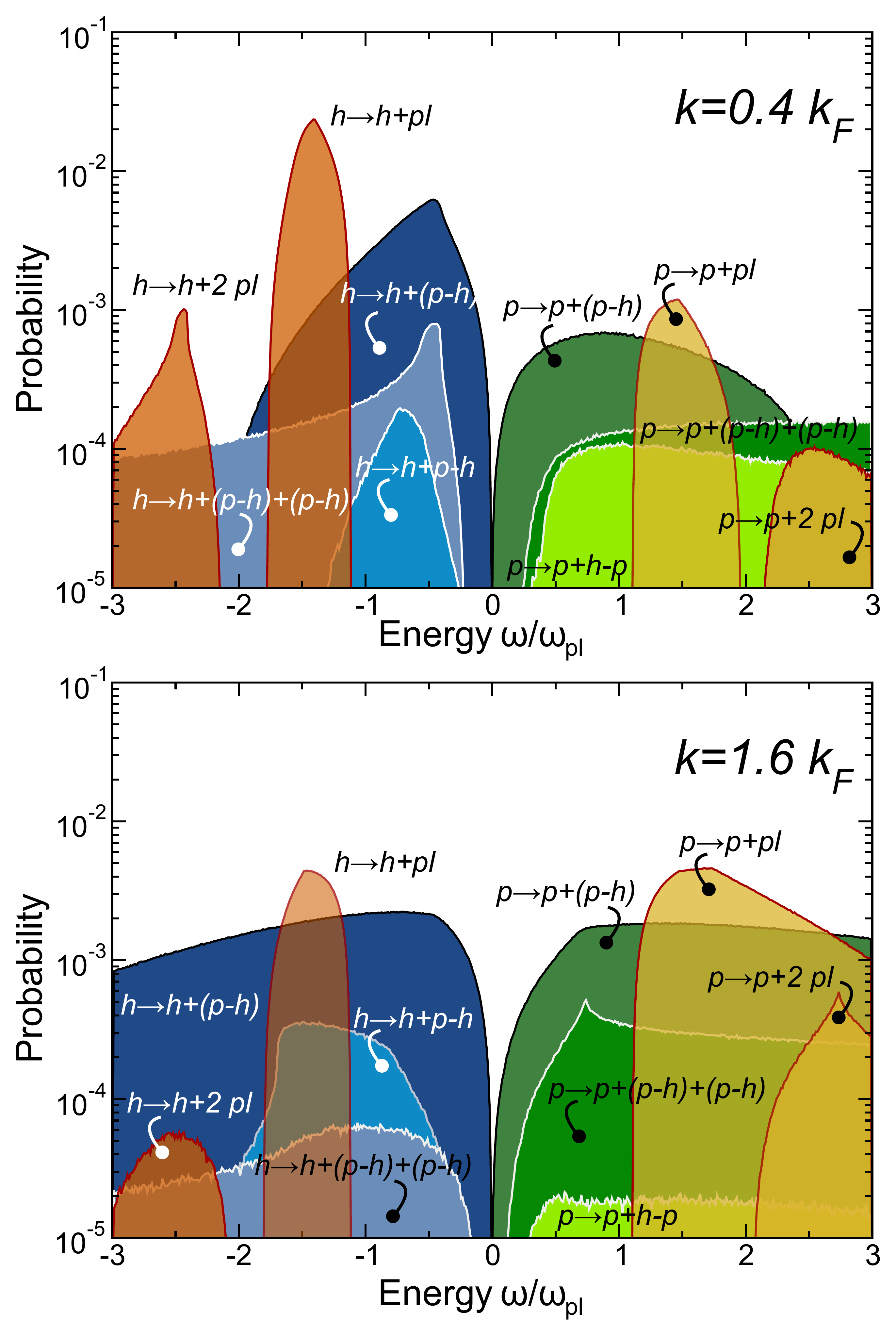}
\caption{Monte-Carlo calculation for 3d HEG at $r_s=5$ density of the first and second
  order diagrams contributing to the electron self-energy (Eq.~\eqref{eq:Dom}). $\mathcal
  D_{-}^{(2b)}$ describes 3 second order processes: generation of two plasmons, of two
  particle-hole pairs, or of one plasmon and one particle-hole pair. The latter as well as
  $\mathcal D_{-}^{(2c)}$ are not shown because they only represent corrections to the
  first-order processes. The Fermi energy is set to zero.
\label{fig:Psgm}}
\end{figure}
\begin{figure}[t!]
\includegraphics[width=\columnwidth]{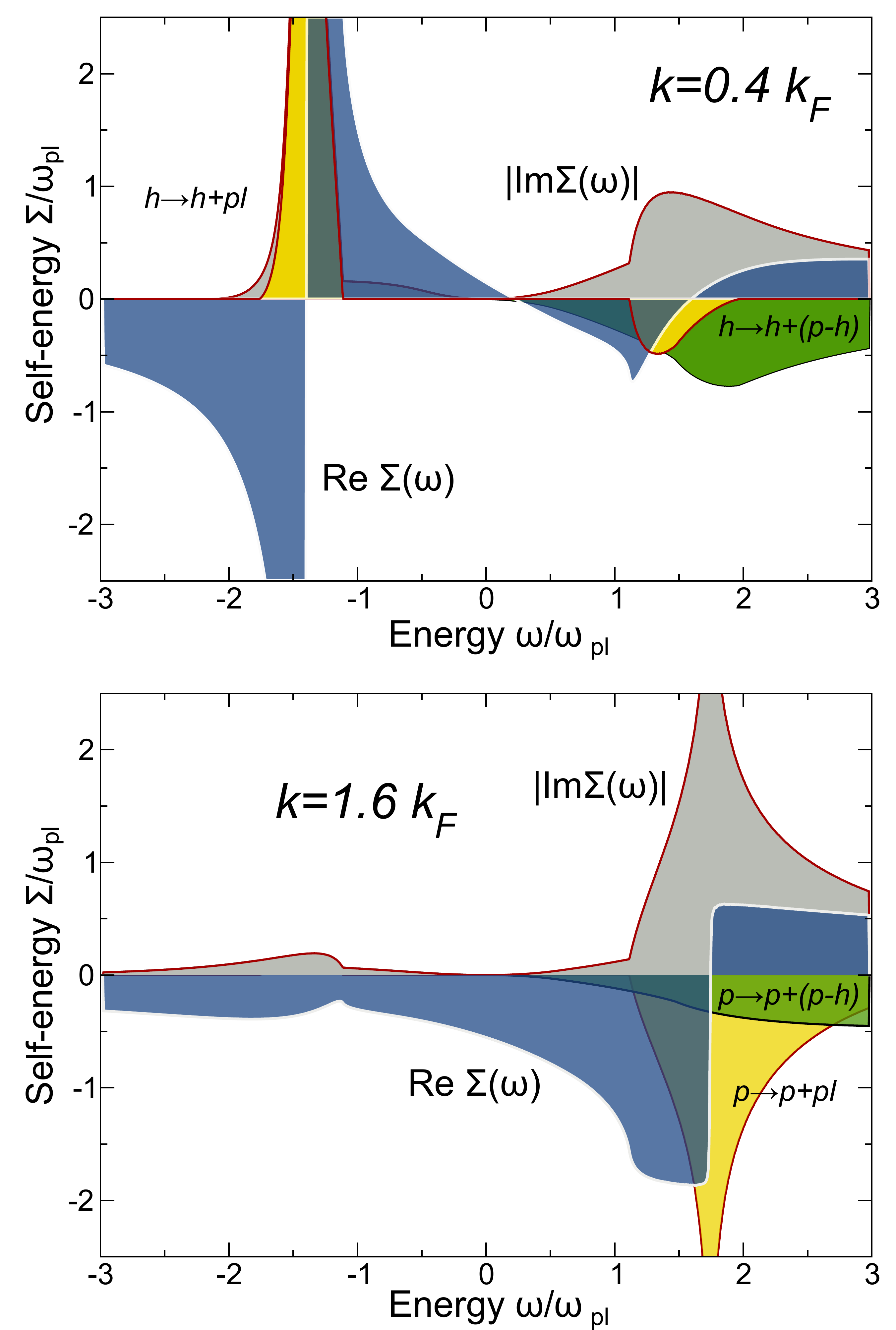}
\caption{Exact first order self-energy of the 3d HEG at $r_s=5$ density. The Fermi energy
  is set to zero. The real part is shown without including the static (exchange) part. The
  values are $e_x=-1.490\omega_{p}$ and $e_x=-0.225\omega_{p}$ for $k=0.4 k_F$ and
  $k=1.6 k_F$, respectively. The plasmon energy in the long wave-length limit is
  $\omega_{p}=2.103\epsilon_F$.
\label{fig:Rsgm}}
\end{figure}

In agreement with our simple argument we see that the phase-space for the first order
processes is limited. The same is observed in the simplest second order process (it
includes also contribution from two bare interaction lines) in view of the same arguments.
The existence of a critical upper momentum for the plasmons also restricts the phase-space
available for the $h\rightarrow h+2pl$ scattering. The situation is completely different
for the $h\rightarrow h+(p\mhyphen h)+(p\mhyphen h)$ events: even though the hole can only
loose a finite momentum the shares between the excitations  can be large
(Fig~\ref{fig:sketch}b), resulting in an arbitrarily large energy transfer
(cf. Eq.~\eqref{eq:D2b}). Hence, the self-energy has an unbounded support, the Paley-Wiener
theorem cannot be applied and the Fermi liquid behavior is restored in the second order.

Our analysis is also important for practical calculations since it allows to determine
\emph{a priori} where a certain diagram might contribute. It is interesting to notice a
sequence of plasmonic peaks in the see of $p\mhyphen h$ excitations. By expanding the
cumulant function $C(k,t)\sim e^{i\omega_p t}$ (Tab.~I, third column) and computing the
Fourier transform one sees that their weight decays as $e^{-a} a^n/n!$. Guzzo \emph{et
al.}~\cite{guzzo_valence_2011} estimated $a\sim 0.3$ for silicon. Therefore, plasmons will
only be important at low orders whereas the tails of the spectral functions are shaped by the
$p\mhyphen h$ scattering mechanisms which lead to the power-law decay. Where such a
crossover occurs depends, of course, on the specific system parameters.

The phase-space arguments provide a partial account of the problem. The inclusion of
matrix elements can modify the self-energy substantially as the comparison of
Fig.~\ref{fig:Psgm} and Fig.~\ref{fig:Rsgm} shows. This can be best seen at the Fermi
level (set to zero in our calculations). While both methods lead to a vanishing
self-energy in this limit the way how it approaches zero is rather different. But how
feasible is the realistic calculation of next order diagrams?  To answer this question let
us consider Eqs.~\eqref{eq:Dom}. There, the second order terms were evaluated by using at
most a 7-dimensional sampling. The full-fledged evaluation, in contrast, would require an
8-dimensional integration for each $k$ and $\omega$ values. In some specific cases
simplifications might be achieved such as in exact analytic treatment of the second-order
exchange term by Onsager \emph{et al.}~\cite{onsager_integrals_1966}. On the other hand,
for practical applications some synthetic approaches might be
promising~\cite{takada_dynamical_2002,bruneval_many-body_2005}.
\section{Conclusions}
In this contribution we performed a detailed analysis of the formula 
$A(t;\epsilon)=A_{QP}(t;\epsilon)\,\exp\left(-\gamma(\epsilon) \frac{t^2}{t+\tau(\epsilon)}\right)$
for the time evolution of the spectral function for extended systems.
The violations such as i)
non-analyticity at short times, ii) reduction of the spectral weight of the quasiparticle
peak or iii) a non-exponential decay at the long-time limit were found. Surprisingly, our
theory reveals that these features are either artifacts of approximations used (iii) or
are rather weak (i) as they result from rather inefficient coupling to $p\mhyphen h$
excitations. We also provide a concise analytic form for the expansion coefficients of the
cumulant function at $t\rightarrow0$.

The short-time limit was analyzed using the cumulant expansion method, while the
asymptotic behavior at longer times was studied using the ordinary many-body perturbation
theory. Thus, it was necessary to establish a connection between both methods. We have
shown that in the first order in the screened Coulomb interaction the cumulant expansion
can be recovered from the MBPT expression for the self-energy, although, some terms are
additionally present in MBPT. For a finite systems we proposed a similar approach and
presented results supported by full numerical calculations for
Na-clusters~\cite{pavlyukh_initial_2011} and C$_{60}$~\cite{pavlyukh_communication:_2011},
which supports the general nature of the time evolution law of the spectral function. The
specific, material-dependent, and quantum size effects are encapsulated in the decay
constants.
\acknowledgments The work is supported by DFG-SFB762 (YP, JB). AR acknowledge financial
support from the European Research Council Advanced Grant DYNamo (ERC-2010-AdG -Proposal
No. 267374), Spanish Grants (FIS2011-65702 C02-01 and PIB2010US-00652), ACI-Promociona
(ACI2009-1036), Grupos Consolidado UPV/EHU del Gobierno Vasco (IT-319-07) and European
Commission projects CRONOS (280879-2 CRONOS CP-FP7) and THEMA(FP7-NMP-2008-SMALL-2,
228539). Computational time was granted by i2basque and BSC Red Espanola de
Supercomputacion.  YP acknowledges enlightening discussions with A. Moskalenko and
M. Sch{\"u}ler on the properties of integrals appearing in Sec.~\ref{sec:limits}.
\appendix
\section{Some momentum integrals\label{A:1}}
A single momentum integral leading involving $\mathcal D_{-}^{(1)}(k,\omega)$ can be
computed as follows:
\begin{equation}
\int\!d\vec q\, f(q,|\vec{k+q}|^2)=\frac{\pi}{k}\int_0^\infty\!q\,dq
\int_{(k+q)^2}^{(k-q)^2}\!dy\,f(q,y),
\label{eq:sph1int}
\end{equation}
with $y=|\vec{k+q}|^2$. The two momenta integrals involing $\mathcal D_{-}^{(2a)}$ can be
computed by introducing symmetrized variables as in~\cite{onsager_integrals_1966}:
\begin{multline}
\int\!d\vec q_1\!\!\int\!d\vec q_2\,
f(|\vec{k-q_1-q_2}|^2,\,|\vec{k-q_1}|^2,\,|\vec{k-q_2}|^2)\\
=\frac18\int\!d\vec q\!\!\int\!d\vec Q\, f(|\vec{k-2q}|^2,\,|\vec{q-Q}|^2,\,|\vec{q+Q}|^2) \\
=\frac{\pi^2}{8k}\int_0^\infty\!dq\int_{(k+2q)^2}^{(k-2q)^2}dy_1
\!\!\int_0^\infty\!dQ\!\int_{qQ}^{-qQ}dy_2\,Q\\
\times f(y_1,\,q^2+Q^2-2y_2,\,q^2+Q^2+2y_2),
\end{multline}
where
\begin{eqnarray*}
\vec q=\vec k-\frac12(\vec q_1+\vec q_2),\quad \vec Q=\frac12(\vec q_1-\vec q_2);\\
y_1=|\vec k-2\vec q|^2,\quad y_2=(\vec q\cdot\vec Q).
\end{eqnarray*}
Analogically for the $\mathcal D_{-}^{(2b)}$ term we have:
\begin{multline}
\int\!d\vec q_1\!\!\int\!d\vec q_2\,
f(|\vec{k-q_1-q_2}|^2,\,|\vec q_1|^2,\,|\vec q_2|^2)\\
=-\frac18\int\!d\vec q\!\!\int\!d\vec Q\, f(|\vec{k-2q}|^2,\,|\vec{q+Q}|^2,\,|\vec{q-Q}|^2) \\
=-\frac{\pi^2}{8k}\int_0^\infty\!dq\int_{(k+2q)^2}^{(k-2q)^2}dy_1
\!\!\int_0^\infty\!dQ\!\int_{qQ}^{-qQ}dy_2\,Q\\
\times f(y_1,\,q^2+Q^2+2y_2,\,q^2+Q^2-2y_2),
\end{multline}
where
\begin{eqnarray*}
\vec q=\frac12(\vec q_1+\vec q_2),\quad \vec Q=\frac12(\vec q_1-\vec q_2);\\
y_1=|\vec k-2\vec q|^2,\quad y_2=(\vec q\cdot\vec Q).
\end{eqnarray*}
\end{document}